\documentclass{article}
\usepackage{epsfig}

\begin{document}

\title{The fitness value of information}

\author{Carl T. Bergstrom\\
  Department of Biology \\
  University of Washington \\
  Seattle, WA 98195-1800, USA\\
  e-mail: {\tt cbergst@u.washington.edu} \and
  Michael Lachmann\\
  Max Planck Inst. for Evol. Anthropology\\
  Deutscher Platz 6\\
  D-04103 Leipzig, Germany\\
  email: {\tt lachmann@eva.mpg.de} }

\maketitle

\bigskip

\begin{abstract}
  Biologists measure information in different ways. Neurobiologists
  and researchers in bioinformatics often measure information using
  information-theoretic measures such as Shannon's entropy or mutual
  information.  Behavioral biologists and evolutionary ecologists more
  commonly use decision-theoretic measures, such the value of
  information, which assess the worth of information to a decision
  maker. Here we show that these two kinds of measures are intimately
  related in the context of biological evolution. We present a simple
  model of evolution in an uncertain environment, and calculate the
  increase in Darwinian fitness that is made possible by information
  about the environmental state. This fitness increase --- the {\em
    fitness value of information} --- is a composite of both Shannon's
  mutual information and the decision-theoretic value of information.
  Furthermore, we show that in certain cases the fitness value of
  responding to a cue is exactly equal to the mutual information
  between the cue and the environment. In general the Shannon entropy
  of the environment, which seemingly fails to take anything about
  organismal fitness into account, nonetheless imposes an upper bound
  on the fitness value of information.
\end{abstract}

\section{Introduction}

Living organisms acquire, store, process, and transmit information ---
and as such, information is a central organizing principle in
biological systems at every scale from the digital coding in DNA to
the long-range calls of cetaceans \cite{MaynardSmith99}. While
information-theoretic measures such as entropy and mutual
information \cite{Wiener48,Shannon48,CoverAndThomas91} have been
embraced in neurobiology and bioinformatics, these measures are less
commonly used in behavioral biology and evolutionary ecology.

The problem is that entropy and mutual information do not directly
address information quality; they do not distinguish between relevant
and irrelevant information. Thus decision theorists, economists, and
behavioral biologists typically measure information by considering its
value: its effect on expected payoff or expected fitness
\cite{Savage54,Good66,Gould74,StephensAndKrebs86,Stephens89,Ramsey90}.

%

\begin{quote}
  {\bf Definition:} The {\em value of information} associated
  with a cue or signal $C$ is defined as the difference between the
  maximum expected payoff or fitness that a decision-maker can obtain
  by conditioning on $C$ and the maximum expected payoff that could be
  obtained without conditioning on $C$.
\end{quote}

The disconnect between entropy and mutual information on one hand and
the value of information on the other has long puzzled biologists in
general and the authors of this paper in particular.  Entropy and
mutual information appear to measure information quantity while
reflecting nothing about fitness consequences; the value of
information measures fitness consequences but has nothing to do with
the actual length or information quantity of a message. But early work
in population genetics
\cite{Haldane57,Kimura61,Felsenstein78,Demetrius83}\footnote{Indeed,
  Claude Shannon wrote a PhD thesis in population genetics before
  embarking on the work that launched the field of information theory
  \cite{Shannon40,Crow01}.} and recent analyses of evolution in
fluctuating environments
\cite{BergstromAndLachmann04,KussellAndLeibler05} hint at a possible
relation between information and fitness. What is this relation?
Information theorists since Kelly \cite{Kelly56} have observed that in
special circumstances, information value and information-theoretic
measures may be related.  Here we argue that these special
circumstances are exactly those about which biologists should be most
concerned: the context of evolution by natural selection. We address
the question ``how much is information worth to living organisms?''
and show that the answer turns out to be a striking amalgam of mutual
information and the decision-theoretic value of information.

\section{A basic model} 
\label{sec:basic}
As evolutionary biologists, how should we measure the cost of
uncertainty\footnote{Numerous studies in population ecology and
  genetics have shown that fitness and population growth in uncertain
  environments depend on the exact nature of the uncertainty; they
  depend both on the distribution of individual reproductive
  successes, and on the correlations in individual successes (reviewed
  in ref.  \cite{YoshimuraAndJansen96}).
  One can capture this complexity by distinguishing between two types
  of uncertainty or risk \cite{RobsonEtAl99}. {\em Idiosyncratic risk}
  is independent of that faced by other individuals, whereas {\em
    aggregate risk} is perfectly correlated among individuals. For
  example, predation imposes largely idiosyncratic risk on a herd of
  herbivores, whereas drought imposes largely aggregate risk. In this
  paper, we focus exclusively on aggregate risk. We will address mixed
  aggregate and idiosyncratic risk in a subsequent report.} or the
value of information? We want to know how the information affects
fitness, so the natural measure of the worth of information is the
following: The {\em fitness value of information}, $G$, is the
greatest fitness decrement or cost that would be favored by natural
selection in exchange for the ability to attain the information.

Like stockbrokers and habitual gamblers, biological organisms faced
with uncertain conditions are selected to behave as if they are
concerned with long-term growth rates.  Thus the fitness value of
information to biological organisms is best measured in terms of the
consequences of this information on the long-term growth rates of
organismal lineages.  Maximizing long-term growth in such conditions
is the same as maximizing the expected value of the logarithm of the
growth rate in a single generation \cite{Cohen66,LewontinAndCohen69} (as
opposed to the expected value of the growth rate itself).

To illustrate these results and to develop an intuition about the
value of information in biological systems, consider the following
simple model of a population of annual organisms living in a variable
environment\footnote{In this section, we follow Cover and Thomas's
  (1991) presentation; these authors offer a parable about a habitual
  gambler who perpetually reinvests his entire winnings at the horse
  track. Their gambling story can be recast quite naturally as a model
  of organisms evolving by natural selection to match their
  physiologies to uncertain environmental conditions.}.  The state of
the environment in each year is an independent random variable $\Phi$ with two
states $\phi_1$ and $\phi_2$, that occur with probability
$p_1$, and $p_2=1-p_1$ respectively. All individuals encounter exactly
the same environment in a given year. 
At the beginning of its development, each organism makes an important
developmental decision to adopt one of two alternative phenotypes: one
suited to environment $\phi_1$, or one suited to $\phi_2$.  The
organism survives to reproduce only if its phenotype properly matches
the demands of the current environment.  The organism's fitness is
given by the following matrix:

\begin{center}  
\begin{tabular}{rcc}
&Phenotype 1&Phenotype 2\\
Environment $\phi_1$&$w_1$&0\\
Environment $\phi_2$&0&$w_2$
\end{tabular}
\end{center}

What should these individuals do in the absence of information about
the condition of the environment? In the short run, individuals
maximize expected fitness by employing the highest-payoff phenotype only.
This yields an expected single-generation fitness of $\max [p_1 \,
w_1, p_2\, w_2]$. 

But in the long run, playing only one strategy will inevitably lead to
a year with zero fitness and consequent extinction. Thus natural
selection will favor not the short run maximization above, but rather
a maximization of long-term fitness. These organisms will be selected
to {\em hedge their bets} during development
\cite{CooperAndKaplan82,SegerAndBrockmann87}, developing into
phenotype 1 with some probability and phenotype 2
otherwise\footnote{Alternatively, organisms can hedge their bets via
  phenotypic switching, as with the bacterial persistence phenotype
  \cite{LachmannAndJablonka96,KussellEtAl05,KussellAndLeibler05}.}.
As we consider a larger and larger span of
generations, natural selection is overwhelmingly likely
\cite{RobsonEtAl99} to favor the strategy that maximizes the growth
rate for 
``typical sequences''
\cite{CoverAndThomas91}, in which
environment $\phi_1$ occurs $N \,p_1$ times, and environment $\phi_2$ occurs $N \,
p_2$ times.  For a genotype that develops with probability $x$ into
phenotype 1, the population growth over such a sequence of $N$ events will be
$(w_1 \,x)^{N p_1}(w_2 (1-x))^{N p_2}$ and 
will be maximized when $N ( p_1 \log( w_1 x ) + p_2 \log( w_2 (1-x) )
)$ is maximized.  This occurs when $p_1 / x = p_2 / (1-x)$ or when
$x=p_1$. Thus for almost all sequences of environments, the strategy
that develops into phenotype 1 with probability $p_1$ will maximize
the growth rate and thus take over the population.  For this strategy,
the expected log growth rate will be $p_1 \log( w_1 p_1) + p_2 \log(
w_2 p_2 )$.

We have set up a simple biological model where uncertainty critically
affects fitness. What is the fitness value of information
here? Suppose that individuals are able to detect a cue that they can
use to forecast the state of the environment with 100\% accuracy. In
this case the organism will use phenotype 1 in environment 1, and
phenotype 2 in environment 2.  What is the fitness value of [being
able to obtain] this cue?

First, we can look at how the cue improves the short-run expected
fitness. With the cue, individuals can always develop the appropriate
phenotype for the environment, and obtain short-run expected fitness
$p_1 w_1 + p_2 w_2$. Thus in the short run, the expected value of
information is $p_1 w_1 + p_2 w_2 - \max [p_1 \, w_1, p_2\, w_2]=\min
[p_1 \, w_1, p_2\, w_2]$. This is exactly the decision-theoretic value
of information.

But natural selection will not maximize short run expected fitness;
instead as discussed above it maximizes the expected log fitness.
Without the cue, the expected log growth rate is $R_{\mbox{\small{no
      inf}}}=p_1 \log (p_1 w_1) + p_2 \log (p_2 w_2)$. With the cue it
is $R_{\mbox{\small{inf}}}=p_1 \log w_1 + p_2 \log w_2$. The fitness
value of information $G$ is the difference between growth with and
without the cue, $R_{\mbox{\small{inf}}}- R_{\mbox{\small{no inf}}}$;
and this quantity is exactly the mutual information between the
perfectly informative cue and the environment, $p_1 \log p_1 + p_2
\log p_2$.  The payoffs $w_i$ have dropped out.  For this very simple
example, the fitness value of information has nothing to do with the
fitnesses $w_1$ and $w_2$, but instead depends exclusively on the
mutual information measure.

This result generalizes naturally to cues that are only partially
informative \cite{Kelly56,CoverAndThomas91}. If the cue is a random
variable $C$, the fitness value of information will be the mutual
information $I(\Phi;C)=\sum_{\phi,c}p(\phi,c) \log {p(\phi,c)\over p(\phi)p(c)} $
between the cue $C$ and the state of the environment $\phi$.

\section{Two illustrative examples} 

Thus far we have been looking at a very special case in which the
fitness of the organism is zero when the wrong phenotype is adopted. A
more realistic model would allow the possibility of non-zero fitness
even when the organism develops to the wrong phenotype.  

\subsection*{Example 1}
We start with a two-environment, two-phenotype example. Since the
players have no control over the state of the environment, we can
study the decision-theoretic behavior of the players without loss of
generality using the following matrix where $1 > a \geq b$:

\begin{center}  
\begin{tabular}{rcc}
&Phenotype 1&Phenotype 2\\
Environment $\phi_1$&1&$b$\\
Environment $\phi_2$&$a$&1
\end{tabular}
\end{center}

If the organism invests $x$ in phenotype 1 and $1-x$ in phenotype 2, her
expected log growth rate will be $p \log[x + a(1 - x)] + (1 - p)\log
[b\, x + (1 - x) ]$. In the absence of information about which
environmental state is realized, the choice of $x^*(p)$ that maximizes
expected log growth given the probability $p$ of environment 1 is:

\begin{equation}
x^*(p) = \left \{
\begin{array}{ll}
0&\textrm{for } p\leq\frac{a(1-b)}{1-a\,b}\\
\frac{p+a\,b(1-p)-a}{(1-a)(1-b)}&\textrm{for }\frac{a(1-b)}{1-a\,b}<p<\frac{1-b}{1-a\,b}\\
1&\textrm{for } \frac{1-b}{1-a\,b}\leq p
\end{array}
\right.
\end{equation}

We see that the organism hedges bets only in a central region. Beyond
that region, the optimal bet-hedging strategy would require the
organism to produce one of the phenotypes with negative probability.
This sort of investment may be feasible in a stock market or a horse
race, but negative bets seem to lack a biological meaning. In
biological situations, we do better to look at the constrained case
where the organism must produce each phenotype with non-negative
probability.

\begin{figure}
\begin{center}
      \includegraphics[width=4 in]{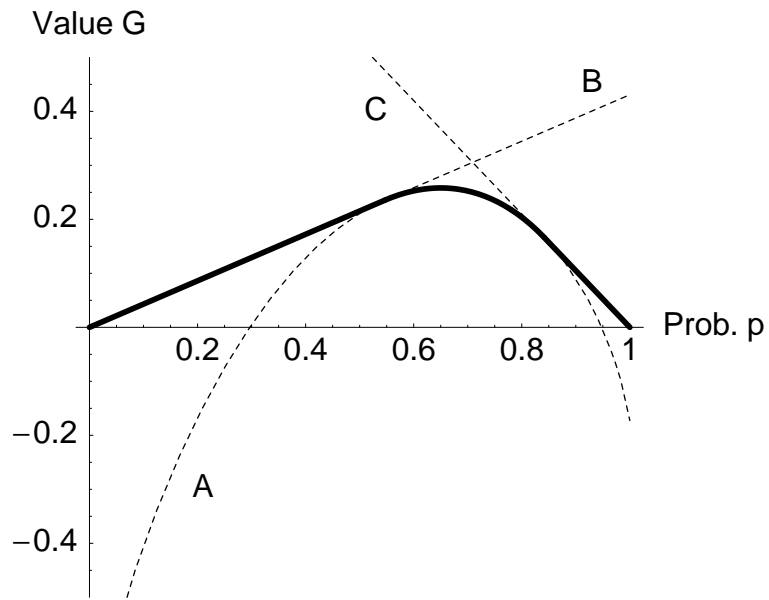}
\end{center}
\caption{ The fitness value of information (heavy solid curve) as a
  function of environmental probabilities $p$ is a composite of three
  value functions: Curve (A) is the sum of mutual information between
  cue and environment and a linear function of the environmental
  probabilities:$ - \Bigl(p \log[p] +(1 - p) \log [1 - p]\Bigr) -
  \Bigl( \log[1 - ab] - (1 - p) \log[1 - a] - p \log[1 - b]\Bigr)$.
  Curves (B) and (C) are the linear functions $- p \log[ a]$ and $ -
  (1-p) \log [ b]$ respectively. Parameter values: $a=0.65$, $b=0.35$.
  Simple calculus reveals that the despite being composed of a linear
  component and a logarithmic component, the fitness value of
  information $G$ is not only continuous but also once continuously
  differentiable.}
 \label{fig:twodimensions}
\end{figure}

\bigskip
\begin{center}[ Figure 1 about here ]\end{center}
\bigskip

If the organism responds to a cue $C$ that gives the exact state of
the environment, she will match her phenotype to the environment
always, for an expected log growth rate of $\log(1)=0$.  The fitness
value of information $G$ is shown in figure \ref{fig:twodimensions}.
In the central region $\frac{a(1-b)}{1-a\,b}<p<\frac{1-b}{1-a\,b}$,
the fitness value of information is equal to the mutual information
$I(\Phi;C)$ between the (perfectly informative) cue $C$ and the
environment $\Phi$, plus a linear function of the probability of each
environment. Outside the range, when the optimal strategy invests in
only one of the phenotypes, the value of the cue is $-p \log[a]$ or
$-(1-p) \log[b]$.  This is simply the decision-theoretic log value of
information, i.e., the expected log of the value one would get if one
took the decision-theoretic approach of maximizing fitness in one
generation.

\subsection*{Example 2}

To get a better intuition of how the fitness value of information
relates to the evolutionarily optimal strategy in the absence of
information, we move to the case of 3 environments that occur with
probabilities $p_1$, $p_2$, and $p_3=(1-p_1-p_2)$. While the principles
generalize to larger numbers of environments and less-symmetric
payoffs, three symmetric environments are far easier to represent
graphically than are the more complicated alternatives.  Thus we
consider the following payoffs structure where $a>1$:

\begin{center}  
\begin{tabular}{rccc}
&Phenotype 1&Phenotype 2&Phenotype 3\\
Env. $\phi_1$&$a$&1&1\\
Env. $\phi_2$&1&$a$&1\\
Env. $\phi_3$&1&1&$a$
\end{tabular}
\end{center}

Using the approach sketched out above, we can compute the fractional
investment ${x_1,x_2,x_3}$ in each strategy that maximizes long-term
growth rate:
\begin{eqnarray}
x_1&=&(p_1(1+a)-p_2-p_3)/(a-1)\nonumber \\
x_2&=&(p_2(1+a)-p_1-p_3)/(a-1)\\
x_3&=&(p_3(1+a)-p_1-p_2)/(a-1)\nonumber
\end{eqnarray}

\begin{figure}
\begin{center}
      \includegraphics[width=4.5 in]{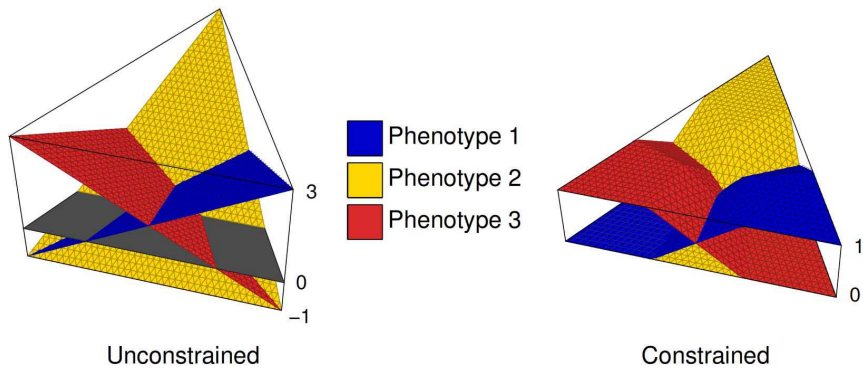}
\end{center}
\caption{ Fractional investment in each strategy in order to maximize
  long-term growth rate, as a function of the probabilities
  $(p_1,p_2,p_3)$ of each environment, for example 2 with $a=2$.
  Because of the constraint $p_1+p_2+p_3=1$, we can represent the
  space of all possible environment probabilities $(p_1,p_2,p_3)$ as
  the two-dimensional simplex where one corner represents $(1,0,0)$,
  another $(0,1,0)$, and the third $(0,0,1)$.  The height of the three
  surfaces at any point indicates the fractional investment in each
  strategy at that point. On the left, we have the unconstrained
  optimum, in which individuals may ``bet against'' certain
  phenotypes, investing negatively in them and putting the suplus into
  the other phenotypes.  On the right is the constrained (and
  biological relevant) solution, in which the fraction invested in
  each phenotype must be non-negative.}
 \label{fig:optbets}
\end{figure}

\bigskip
\begin{center}[ Figure 2 about here ]\end{center}
\bigskip

This optimal strategy is shown in the left panel of Figure
\ref{fig:optbets}. Here we have a curious sort of investment; the gray
surface is the ``invest zero'' plane. When the colored surfaces drop
below this, the player is effectively betting against those phenotypes
by producing them with negative probability --- which makes no
biological sense, as discussed above. Our solution is then only
reasonable in the central region where all three bets are
non-negative. This area, which we will call Region 1, is delimited by
$p_i>1/(2+a)$ for all $i=1,2,3$.  Outside of Region 1, we will have to
compute optimal bets subject to constraints that no bet is negative.
We do this below and illustrate the result in the right panel of Figure
\ref{fig:optbets}.

When one environment is sufficiently infrequent but the other two are
common, an individual will invest in the phenotypes corresponding to
the two common environments but not in the rare one. There are three
such regions on the simplex, with boundaries given by the trio of inequalities
$p_i<1/(2+a)$, $p_j<p_k a$, and $p_k<p_j a$.  In these three areas
which collectively we call Region 2, optimal allocation is given by
\begin{eqnarray}
x_i&=&0 \nonumber \\
x_j&=&(p_j a-p_k) / (a-1)\\
x_k&=&(p_k a-p_j)/(a-1)\nonumber
\end{eqnarray}
Finally, when two environments are sufficiently rare, individuals will
produce only the phenotype corresponding to the common environment.
This occurs outside of the areas covered by Regions 1 and 2, in three
corner areas which collectively we call Region 3.

Because of the different betting strategies in each region, the value
of information in each region is computed by a different formula. We
take these in turn. In Region 1, a cue indicating the state
of the environment increases the expected log growth rate by
\begin{equation}
\log [a / (2+a)] - \sum p_i \log p_i\, =\, \log [a / (2+a)] + I(\Phi,C)
\end{equation}
This is simply a constant plus the mutual information between
the environment and the (perfectly informative) cue.

In Region 2, let $l$ be the phenotype never adopted by the organism.
Then the cue increases the expected growth rate by
\begin{equation}
\log a - \sum_{i \neq l} p_i \log p_i - (1-p_l)\log[(1+a)/(1-p_l)]
\end{equation}

In Region 3, let $l$ be the phenotype always adopted by the
organism. The cue increases the expected log growth rate by
\begin{equation}
 (1-p_l)\log a
\end{equation}
This is simply the decision-theoretic log value of information, i.e.,
the log of the value one would get if one took the decision-theoretic approach of
maximizing fitness in one generation.

\begin{figure}
\begin{center}
 \includegraphics[width=4.5 in]{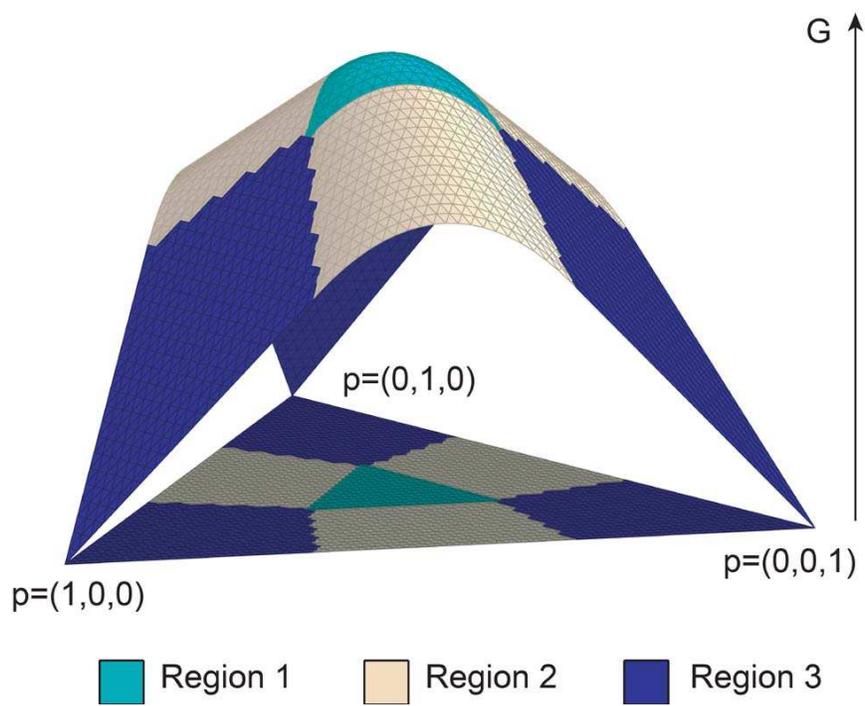}\end{center}

\caption{ The fitness value of information $G$ as a
function of the probabilities that each environment occurs, for the
symmetric three-environment scenario with $a=2$, displayed on the
simplex $p_1+p_2+p_3=1$.}
 \label{fig:Gsurface}
\end{figure}

Putting these all together, we get the surface shown in Figure \ref{fig:Gsurface}.

\bigskip
\begin{center}[ Figure 3 about here ]\end{center}
\bigskip

Surprisingly, this fitness value of information surface seamlessly
sews together a region described by the mutual information (Region
1), a region described by the decision-theoretic value of information
(Region 3), and an intermediate region (Region 2). Comparing the
height of the surface and the gradients along the relevant edge and
point boundaries, calculus reveals that this surface is again
continuous and once continuously differentiable 
everywhere. The fitness value of information incorporates both the
information-theoretic measure and the decision-theoretic value --- and
through the continuity of the corresponding regions, we also see a
fundamental connection between these two measures.

\section{Extending the model}
\label{sec:mainmodel}  
Let us now assume that an organism has to make a developmental
decision between $n$ possible phenotypes, each of which is
a best match to one of $n$ environments. The environments $\phi_i$ occur with
probabilities $p_i$ and the fitness of phenotype $j$ in environment
$i$ is $w_{ij}$.

How should an organism respond? To maximize short-run expected fitness,
the organism should simply develop the phenotype with the highest
expected fitness. Expected fitness then will be $E[w]=\max_j \Bigl[
\sum_i p_i\,w_{ij} \Bigr]$.

What about long-term fitness?  We can find a general form for the
fitness value of information for those cases corresponding to Region 1
in the previous example, i.e., where the organism develops into all
$n$ phenotypes with positive probability. Let us look at a
strategy that produces phenotype $i$ with frequency $x_i >  0$.
The organism will be selected to maximize the expected log growth rate
$R$, so we want to find the strategy that maximizes the log growth
rate $\sum_i p_i \log \sum_j w_{ij} x_j$ subject to the constraint
that the fractional investments in the various phenotypes sum to one:
$\sum_i x_i=1$.  The Lagrangian for this problem is

\begin{equation}
L(x_1,x2_,\dots,x_n,r)=\sum_i p_i \log
\sum_j w_{ij} x_j  - \lambda \Bigl(\sum_i x_i - 1 \Bigr)
\end{equation}

Since the constraint function is a linear function, it immediately
satisfies the constraint qualification that the partials of the
constraint function at the constrained maximizer are not all zero. We
maximize the Lagrangian by taking partial derivatives and setting to
zero. The partials with respect to $x_k$ yield a set of $n$ equations:
\begin{equation}
  \label{eq:growth_rate_n}
  \frac{\partial}{\partial x_k} \bigl[ \sum_i p_i \log \sum_j w_{ij} x_j
  \bigr. - \lambda \sum_i x_i] = 0
\end{equation}
Assuming that $W$ (the matrix whose $(i,j)$ entry is $w_{ij}$) is
invertible, we can write $y_i = \sum_j w_{ij} x_j$, and $V=W^{-1}$, so
that $x_j = \sum_i v_{ji} y_i$. Then we can solve
\begin{equation}
  \frac{\partial}{\partial y_k} \bigl[ \sum_i p_i \log \sum_j y_j
   - \lambda \sum_{ij} v_{ji} y_i \bigr] = 0
\end{equation}
for all $k$, which gives for all $k$:
\begin{equation}
  \frac{p_k}{y_k}-\lambda \sum_j v_{jk} = 0
\end{equation}

Now we can solve for the constraint $\sum_i x_i=1$, which gives 
$\lambda=1$ and thus
we have $y_k=p_k / \sum_j v_{jk}$. Though this solution for $y_k$ always exists, and the corresponding $x_i$ always satisfy the constraint $\sum_i x_i = 1$, a solution might contain negative $x_i$, which would not be biologically plausible. Thus the rest of the derivation will assume that we are in a region of parameter space in which the solution is indeed non-negative for all $x_i$. Substituting $y_k$ into the
equation for the log growth rate, to get the maximal log growth rate without the cue,
gives

\begin{equation}
\label{eqn:unconstrained}
R_{\mbox{\small{no inf}}}=\sum_i p_i \log (\frac{p_i}{\sum_j v_{ji}}) = \sum_i p_i \log(p_i) - \sum_i p_i \log( \sum_j v_{ji})
\end{equation}

\noindent This expression is simply $-H(\Phi)+L(p)$, where $H(\Phi)$ is the entropy of the environment and $L$ is a linear function of the probabilities of each environmental state.

We would now like to calculate the value of a cue. This will be the difference in the expected log growth rate between this optimal strategy, and the optimal strategy when a cue $C$ is received. 

First we consider a cue that reveals the exact environment. The organism 
will maximize fitness by matching phenotype to the known environment,
yielding a log growth rate
$R_{\mbox{\small{inf}}}=\sum_i p_i
    \log(w_{ii})$. Thus the value of this cue is
\begin{equation}
\label{eqn:valuemiddle}
   J= R_{\mbox{\small{inf}}}-R_{\mbox{\small{no inf}}}=-\sum_i p_i \log(p_i) + \sum_i p_i \log(w_{ii} \sum_j v_{ji})
\end{equation}

This is the mutual information between the (perfectly
informative) cue and environment, $I(\Phi;C)=H(p)-0$ plus a linear function of the probabilities $p_i$.

Next we assume that the cue does not reveal the exact state of the
environment, but instead only contains partial information about the
environment. Let the mutual information between the cue $C$ and the
environment be $I(\Phi;C)$.  The strategy of the organism will depend
on the cue. We can thus maximize growth rate for each cue separately
using the conditional probabilities of all environments
$\mbox{Pr}(\phi |c)$, and the same argument we used above to calculate
the optimal strategy. We used two assumptions there: first, that $W$
is invertible, and this still holds, and second that in our solution
all $x_k$ are non-negative. We now assume that this is true for the
responses to all cues, and limit the domain of our solution
correspondingly, as discussed below. We can then compute the maximal
growth rate by averaging equation \ref{eqn:unconstrained} over all
cues, for a maximal growth rate of:

 
\begin{eqnarray}
\label{eqn:conditional}
R_{\mbox{\small{inf}}}=\sum_c \mbox{Pr}(c) \sum_\phi \mbox{Pr}(\phi |c) \log(\mbox{Pr}(\phi |c)) - \sum_c \mbox{Pr}(c) \sum_\phi \mbox{Pr}(\phi |c) \log( \sum_j v_{ji}) \nonumber \\
= \sum_c \sum_\phi \mbox{Pr}(\phi,c) \log(\mbox{Pr}(\phi |c)) - \sum_i p_i \log( \sum_j v_{ji})
\end{eqnarray}

This is $-H(\Phi |C)+L(p)$, where $H(\Phi|C)$ is the conditional
entropy of the environment given the cue. The fitness value of
information $G$ conferred by the cue is the difference in the growth
rates: $-H(\Phi | C)+L(p)+H(\Phi)-L(p)=I(\Phi;C)$. Thus $G$ is simply the mutual
information between the cue and the environment.

Note that the fitness difference will be exactly $I(\Phi;C)$ only when
the organism produces all phenotypes with positive probability both
with and without the cue. In our calculation we assumed that for all
cues, all $x_i$ are positive. If the environmental probabilities are
such that the organism hedges in the absence of a cue, then if the cue
conveys sufficiently little information, the organism will also do
best to hedge after receiving the signal as well, albeit with
different fractions going into each phenotype.  The reason is that for
each particular signal $c$, we get a solution using conditional
probabilities $\mbox{Pr}(\phi |c)$ instead of the original
probabilities $\mbox{Pr}(\phi)$. If all the $\mbox{Pr}(\phi |c)$ are
sufficiently close to $\mbox{Pr}(\phi)$, they will fulfill the same
requirements that $\mbox{Pr}(\phi)$ fulfills, and thus there will be a
solution with positive $x_i$ for each signal 
In all such cases, the fitness gained from a signal will be exactly
the mutual information between the signal and the environment. When we
are outside this range, the gain from a signal will be lower relative
to the mutual information, and can even be 0. For example, if without
the signal no bet-hedging occurs, and all signals convey so little
information that no decision is changed, then the gain in growth
rate resulting from the signal will be 0.

\section{Bounding the fitness value of information}

We can also show that the fitness value of information is bounded
above by the mutual information between cue and environment.
Compare the expected log growth rate of individuals of two
types.  Type A individuals receive a cue $C$ with possible values
 ${c_1,c_2,c_3,\dots,c_n}$ drawn from a distribution with probability
function $\mbox{Pr}(C)$ and entropy $H(C)$.  Each individual then maximizes
expected log growth rate by following some investment strategy $s(c)$
that sets how to invest in the various phenotypes, given the receipt
of cue $c$. 

Type B individuals do not receive this cue. Instead, they follow the
betting strategy $r=\sum_c \mbox{Pr}(c)s(c)$, thereby employing a
probability-matching mixture of the various $s(c)$ strategies used by
Type A individuals.

Represent the fitness of an individual using strategy $s(y)$ when the
cue was $c$ by $w(s(y)|c)$.  The expected log growth rate for Type A
individuals is then 

\begin{equation}R_A=\sum_c \mbox{Pr}(c) \log [ w(s(c)|c) ].
\end{equation}

The expected log growth rate for Type B individuals is
\begin{equation}R_B=\sum_c \mbox{Pr}(c) \log \Bigl[ \sum_{c'}\mbox{Pr}(c') w(s(c')|c)\Bigr].\end{equation}

Since fitnesses are non-negative, the $w(s(c')|c)$ terms in the
summation above must be at least zero even for $c' \neq c$, and
therefore $\sum_c' \mbox{Pr}(c') w(s(c')|c) \geq \mbox{Pr}(c) w(s(c)|c)$. Since $\log$ is a
monotone function, this implies $\log \Bigl[\sum_c' \mbox{Pr}(c') w(s(c')|c)
\Bigr] \geq \log \Bigl[ \mbox{Pr}(c) w(s(c)|c) \Bigr].$ Thus:

\begin{eqnarray}R_B&\geq& \sum_c \mbox{Pr}(c) \log [ \mbox{Pr}(c)
w(s(c)|c)]\nonumber \\&=&\sum_c \mbox{Pr}(c) \log [ w(s(c)|c) ] + \sum_c \mbox{Pr}(c) \log \mbox{Pr}(c)\nonumber\\& =&
R_A - H(C).
\label{eqn:lambdaB}
\end{eqnarray}

Since a growth rate of at least $R_B$ can be attained without
information, the fitness value
of information is therefore bounded by $G\leq R_A-R_B<H(C)=I(\Phi,C)$ Thus the fitness
value of information is at most equal to the mutual information
between a perfectly informative cue and the environment, {\em
  irrespective of the actual fitness payoffs $w$.}  As before, this
result can be generalized to partially informative cues
\cite{CoverAndThomas91}. 

\section{Discussion}

In this paper we have shown that two measures of information, the
information-theoretic mutual information and the decision-theoretic
value of information, are united into a single measure when one looks
at the strategies that natural selection will favor, namely those that
maximize the long term growth rate of biological organisms.
Furthermore, we have shown that under conditions in which bet-hedging
is advantageous, and with cues that convey little information, the
fitness value of information associated with those cues is exactly the
mutual information between the cue and the environment.  Finally, we
have shown that the fitness value of an informative cue is bounded
above by the mutual information between that cue and the environment,
and in some cases is equal to exactly this value.  These results
establish a close relationship between biological fitness and
information-theoretic measures such as entropy or mutual information.

But why does this relation exist? To answer that question, we should
take a closer look at the concept of information: information is the
reduction of uncertainty, where uncertainty measures the number of
states a system might be in.  Thus mutual information between the
world and a cue is the fold reduction in uncertainly about the world after
the cue is received.  For example, if a system could be in any of six
equiprobable states, and a cue serves to narrow the realm of
possibility to just three of these, the cue provides a twofold
reduction in uncertainty.  For reasons of convenience, information
is measured as the logarithm of the fold reduction in uncertainty ---
this ensures that the measure is additive, so that for example we can
add the information received by two successive cues to calculate the
total information gained \cite{Nyquist24,Hartley28,Shannon48}.

Thus while information concepts are often thought to be linked with
the famous sum $\sum p \log (p)$, the fundamental concept is not a
particular mathematical formula. Rather, it is the notion that
information measures the fold reduction in uncertainty about the possible
states of the world. 


With this view, it is easy to see why information bears a close
relation to biological fitness: For simplicity, consider an extreme
example in which individuals survive only if their phenotype matches
the environment exactly, and suppose that there are ten possible
environments that occur with equal probability. In the absence of any
cue about the environment, the best the organism can do is randomly
choose one of the ten possible phenotypes with equal probability. Only
one tenth of the individuals will then survive, since only a tenth
will match the environment with their phenotype. If a cue conveys 1
bit of information and thus reduces the uncertainty about the
environment twofold, the environment can be only in one of five possible
states. The organism will now choose randomly one of five possible
phenotypes, and now a fifth of the population will survive --- a
twofold increase in fitness, or a gain of 1 bit in the log of the
growth rate. 
%

What happens when the environments are not equiprobable? In this case
we can understand the connection between information and fitness by
looking to long sequences of environments and the theory of typical
sequences . The theory tells us that almost surely one of the
``typical sequences'' --- those sequences in which the environments
occur in their expected frequencies --- will occur
\cite{CoverAndThomas91}.  Moreover, all typical sequences occur with
equal probability. Thus a lineage is selected to divide its members
equally among all typical sequences.  Since any one mistake in
phenotype is lethal, only a fraction of these lineages, those that
have just the right sequence, will survive. The number of typical
sequences in this case is exactly $2^{N H(\Phi)}$ where $N$ is the
number of generations in the sequence and $H(\Phi)$ is the entropy of
the environment.  Correspondingly, the fraction of surviving lineages
will be $2^{-N H(\Phi)}$ . If a cue $C$ is received that reduces the
uncertainty of the environments by $I(\Phi;C)$, then the fraction of
surviving lineages can be increased by exactly $2^{N I(\Phi;C)}$. This
is analogous to the situation in communication: if we need to encode a
string of symbols that are not equiprobable, we turn to a long
sequence of such symbols. Our code then needs only to be efficient for
representing typical sequences of symbols, and those typical sequences
occur with equal probability. The number of such sequences is $2^{N
  H}$, where $N$ is the length and $H$ is the entropy of the symbols.
If the message recipient also obtains side information related to the
message itself, then the mutual information
$I(\mbox{message};\mbox{side information})$ measures the reduction in
the number of possible messages that need to be encoded by the
transmitter. This number of messages is reduced by exactly $2^{N
  I(\mbox{message};\mbox{side information})}$-fold by the presence of
the side information.

We can now see why the concept of information is the same across
different disciplines. In biology, fitness refers to the fold increase
in the number of surviving lineages. In communication theory,
information refers to the fold increase in the number of messages to
encode. In physics, entropy refers to the fold increase in the number
of possible states in phase space.

Finally, our results also suggest that information theory will be
useful in studying the evolution of communication.  Even before
knowing what a biological signal means, how it is used, or what the
fitness structure of the environment may be, we have shown that one
can place an upper bound on the fitness consequences of responding to
that signal, simply by measuring the information content of the
signalling channel.


\newpage

\section*{Acknowledgments}

This work was supported in part by a UW RRF award to CB and was
initiated during the authors' visit to the H. R. Whiteley Center in
Friday Harbor, WA.  The authors thank Sidney Frankel, Matina Donaldson
and Arthur Robson for their helpful discussions, and David Park for
his help in producing the figures. ML completed the early stages of
this paper while at the Max Planck Institute for Mathematics, which
provided a very fruitful environment to work on such questions.

\bibliography{sig,language,information,econ,decision,fluct,bergstrom,ESS,sex}

\bibliographystyle{unsrt}

\end{document}